\documentclass[twocolumn,superscriptaddress,aps,showpacs,amsmath,footinbib,bibnotes]{revtex4-1}
\pdfoutput=1
\usepackage{graphicx}
\usepackage{amssymb,amsmath}
\usepackage{textcomp}
\usepackage[bold]{hhtensor}
\usepackage{natbib}
\usepackage{lipsum}
\usepackage{bm}
\usepackage{hyperref}
\usepackage{multirow}
\usepackage{float}
\usepackage{amsmath}
\usepackage{amssymb}
\usepackage{stmaryrd}
\usepackage{mathrsfs}
\usepackage[dvipsnames]{xcolor}
\hypersetup{
	colorlinks,
	linkcolor={red!80!black},
	citecolor={Blue},
	urlcolor={MidnightBlue}
}

\begin{document}

\author{Tianhao Ren}
\affiliation{John A. Paulson School of Engineering and Applied Sciences, Harvard University, Cambridge, Massachusetts 02138, USA}
\affiliation{School of Electronic Science and Engineering, University of Electronic Science and Technology of China, Chengdu, 611731, China}
\author{Mian Zhang}\email{mian@hyperlightcorp.com}
\affiliation{John A. Paulson School of Engineering and Applied Sciences, Harvard University, Cambridge, Massachusetts 02138, USA}
\affiliation{HyperLight Corporation, 501 Massachusetts Avenue, Cambridge, MA, 02139, USA}
\author{Cheng Wang}
\affiliation{John A. Paulson School of Engineering and Applied Sciences, Harvard University, Cambridge, Massachusetts 02138, USA}
\affiliation{Department of Electronic Engineering \& State Key Lab of THz and Millimeter Waves, City University of Hong Kong, Kowloon, Hong Kong, China}
\author{Linbo Shao}
\affiliation{John A. Paulson School of Engineering and Applied Sciences, Harvard University, Cambridge, Massachusetts 02138, USA}
\author{Christian Reimer}
\affiliation{John A. Paulson School of Engineering and Applied Sciences, Harvard University, Cambridge, Massachusetts 02138, USA}
\affiliation{HyperLight Corporation, 501 Massachusetts Avenue, Cambridge, MA, 02139, USA}
\author{Yong Zhang}
\affiliation{John A. Paulson School of Engineering and Applied Sciences, Harvard University, Cambridge, Massachusetts 02138, USA}
\affiliation{School of Electronic Science and Engineering, University of Electronic Science and Technology of China, Chengdu, 611731, China}
\author{Oliver King}
\affiliation{Collins Aerospace, Advanced Technology, Cedar Rapids, IA, 52402, USA}
\author{Ronald Esman}
\affiliation{Collins Aerospace, Advanced Technology, Cedar Rapids, IA, 52402, USA}
\author{Thomas Cullen}
\affiliation{Collins Aerospace, Advanced Technology, Cedar Rapids, IA, 52402, USA}
\author{Marko Loncar}\email{loncar@seas.harvard.edu}
\affiliation{John A. Paulson School of Engineering and Applied Sciences, Harvard University, Cambridge, Massachusetts 02138, USA}
\date{\today}
\title{An integrated low-voltage broadband lithium niobate phase modulator}

\begin{abstract}
Electro-optic phase modulators are critical components in modern communication, microwave photonic, and quantum photonic systems. Important for these applications is to achieve modulators with low half-wave voltage at high frequencies. Here we demonstrate an integrated phase modulator, based on a thin-film lithium niobate platform, that simultaneously features small on-chip loss ($\sim$ 1 dB) and low half-wave voltage over a large spectral range (3.5 - 4.5 V at 5 - 40 GHz). By driving the modulator with a strong 30-GHz microwave signal corresponding to around four half-wave voltages, we generate an optical frequency comb consisting of over 40 sidebands spanning 10 nm in the telecom L-band. The high electro-optic performance combined with the high RF power-handling ability (3.1 W) of our integrated phase modulator are crucial for future photonics and microwave systems.
\end{abstract}
\maketitle

Electro-optic (EO) phase modulators are fundamental building blocks in photonics and microwave systems. Possible applications include photonic radio-frequency (RF) front ends for radio-over-fiber systems \cite{clark_photonics_2011}, telecommunication systems \cite{janner_microstructured_2009,miller_attojoule_2017,wang_integrated_2018,wooten_review_2000}, EO frequency combs \cite{zhang_broadband_2018}, temporal imaging systems \cite{kolner_space-time_1994} and quantum photonics \cite{kues_-chip_2017}. Among various material platforms explored, including Si \cite{dong_low_2009}, InP \cite{li_high_2010}, AlN \cite{xiong_aluminum_2012}, plasmonics  \cite{melikyan_high-speed_2014}, graphene \cite{mohsin_experimental_2015} and polymers \cite{lee_broadband_2002}, lithium niobate (LN) has been the most widely used for high-end phase modulator applications. This is due to LN’s outstanding material properties, including good temperature stability, wide transparency window and most importantly highly efficient linear EO (Pockels) effect \cite{wooten_review_2000,arizmendi_photonic_2004,boes_status_2018}. Commercially available LN phase modulators today are generally based on bulk LN crystals, where a weak optical index contrast is created by ion in-diffusion or proton exchange technologies \cite{bazzan_optical_2015}. The low index contrast results in low optical confinement and high half-wave voltage (V$_\pi$), the latter of which is defined as the voltage required to switch the phase of light by 180$^\circ$. For example, the V$_\pi$ of a typical commercial phase modulator operating at 1525 - 1605 nm is 7.5 V at 30 GHz \cite{thorlabs_40_nodate}.

Thin-film lithium niobate on insulator (LNOI) platform \cite{boes_status_2018,poberaj_lithium_2012} has recently emerged as a promising approach to realize integrated EO modulators with stronger optical confinement and high EO efficiencies while occupying a smaller footprint \cite{wang_integrated_2018,cai_electric-optical_2016,stenger_single-sideband_2018,rao_high-performance_2016,jin_linbo_2016,macario_full_2012,mercante_thin_2016,weigel_hybrid_2018}. For example, LNOI phase modulators with relatively low V$_\pi$ (3.8 V at DC) have been demonstrated \cite{mercante_thin_2016}, albeit at the expense of a high on-chip optical loss of $\sim$ 10.5 dB.

Here we demonstrate a dual channel phase modulator (Fig. 1) using a thin-film LNOI platform with low on-chip optical loss ($\sim$ 1 dB/channel) and low RF V$_\pi$ (4.5 V at 40 GHz) that is relatively constant (less than 28\% variation) over wide frequency range (from 5 to 40 GHz). The low RF V$_\pi$, high RF power-handling capability and low optical losses together allow us to generate an EO frequency comb with over 40 sidebands using a single modulator, instead of cascading multiple devices \cite{papp_microresonator_2014}.

We first fabricate the optical waveguide using electron-beam lithography (EBL) followed by reactive-ion etching with argon (Ar+) ions \cite{zhang_monolithic_2017,wang_integrated_2018}. Next, gold RF electrodes with spatial separation of 5 $\mu$m are placed along the waveguides. In contrast to our previous work \cite{wang_integrated_2018}, where a second EBL was used to define the electrodes, here we use direct-write laser lithography (Heidelberg MLA150). This allows us to pattern the electrodes more than 30 times faster than EBL, while maintaining a good alignment accuracy. 

\begin{figure}
	\centering
	\includegraphics[angle=270,width=0.5\textwidth]{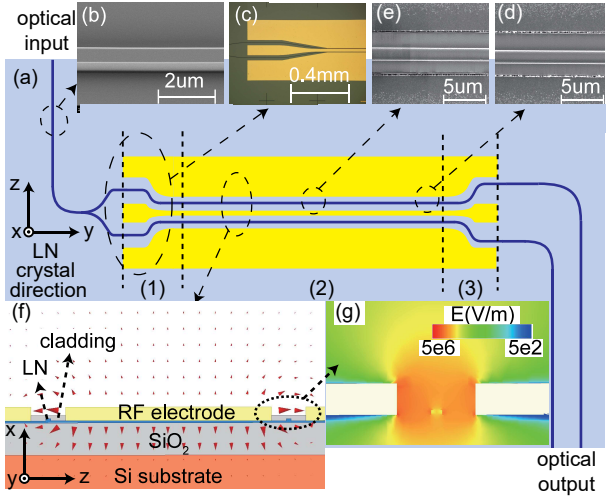}
	
	\caption{\label{fig1}\textbf{Device geometry} (a) Schematic (top view) of the phase modulator. Sections (1) and (3) are used as the input and output contact pads for RF probes, while (2) is the modulation section; (b) SEM image of the optical waveguide before cladding; (c) Microscope image of device showing the input section of the RF electrodes; (d) and (e) SEM images showing the alignments between waveguides and electrode at different positions; (f) Numerically simulated RF electric field distribution and (g) electric field intensity around the gap area in the cross-section view of the phase modulator. The coordinates in (a) and (f) indicate the LN crystal direction.}
\end{figure}

In order to achieve low RF V$_\pi$ values and high optical bandwidths, a 50-$\Omega$ impedance-matched traveling-wave RF transmission line is used, where the optical and RF group velocities are matched \cite{wang_integrated_2018}. To achieve this, we employ coplanar waveguide geometry with ground-signal-ground (GSG) configuration, where the electric fields in the two gaps have opposite directions (Fig. \ref{fig1}). The input and output sections are designed to have wider signal lines for RF probes contacting purposes, since the 30-$\mu$m-wide signal line in the modulation section is too narrow for electrical probes. The optical signal is equally split into two paths, which are modulated separately. The push-pull phase modulator design is of interest for fiber-optic gyroscopes where differential phase shift can be useful \cite{huang_brillouin_1993} and phase-modulated optical links \cite{urick_fundamentals_2015}. For other applications the Y-splitter can be removed to reduce the associated 3 dB loss. Our numerical modeling indicates that the majority of the electric field is focused around the gap area (Fig. \ref{fig1}(f) and Fig. \ref{fig1}(g)), indicating  a good overlap between optical and microwave fields. The microwave electrodes are placed such that the major field component is aligned with the $z$ direction of our x-cut LN crystal (Fig. \ref{fig1})  in order to make use of the strongest EO coefficient $r_{33}$ \cite{wooten_review_2000}.  The strong optical confinement in our structures (index contrast $\sim$ 0.67) allows for a small electrode-waveguide gap (5 $\mu$m) resulting in a high EO modulation efficiency. The rib waveguide has a top width of 700 nm, which allows single-mode operation for transverse-electric (TE) polarization \cite{wang_nanophotonic_2018}. The total thickness of the LN layer is 600 nm, with top 300 nm etched to form the rib waveguide. The high-quality etching is confirmed by a scanning electron microscope (SEM) examination before cladding (Fig. \ref{fig1}(b)), showing smooth sidewalls of the optical waveguide.

Longer metal electrodes are preferred for minimizing the driving power (and voltage) since they can induce more phase shift at a given driving voltage. In practice, however, longer electrodes result in higher microwave losses, which reduces the bandwidth of the phase modulator. Moreover, minor misalignment in the electrode fabrication (visible in Fig. \ref{fig1}(d) and Fig. \ref{fig1}(e)) can result in optical absorption due to adjacent metal electrodes. Here, we choose a trade-off with the modulation section length being 2 cm long (Fig. \ref{fig1}(a)). To reduce the microwave transmission losses, electrodes thickness should be larger than the skin depth at the frequency of interest. We fabricate 1.6-$\mu$m thick microwave electrodes, using a gold evaporation process, which is about 1.5 times the skin depth at 5 GHz and about 4 times the skin depth at 40 GHz.

\begin{figure}
	\centering
	\includegraphics[angle=0,width=0.45\textwidth]{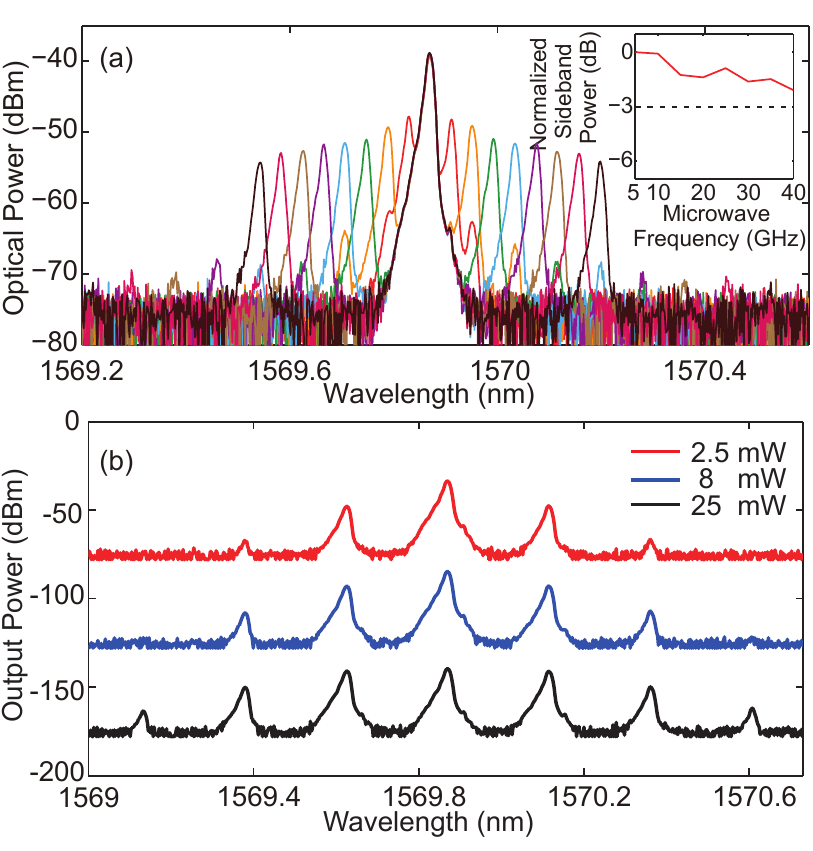}
	
	\caption{\label{fig2}\textbf{Inverse taper mode matching.} (a) Optical spectra of the phase-modulated signals when the modulator is driven by a RF signal of  8 mW at different frequencies (from 5 to 40 GHz, in steps of 5 GHz). The inset shows the normalized powers of the first sidebands as a function of frequency, after subtracting the RF cable losses, showing a roll-off of 2.1 dB from 5 to 40 GHz. (b) Optical spectra of the phase-modulated signals with increasing RF powers of 2.5 mW (red line), 8 mW (blue line), and 25 mW (black line) at 30 GHz. Each offset is 50 dBm.}
	
\end{figure}

To characterize the modulators, light from a tunable telecom laser (Santec TSL-510) is launched into the device using a lensed fiber (OZoptics). The transmitted light is collected using another lensed fiber and detected using an InGaAs photodetector (New focus 1811). The overall transmission loss from the input fiber to one of the output fibers is measured to be 16.9 dB, with 3 dB originating from the Y-splitter. The remaining 13.9 dB loss is largely due to input and output coupling losses at chip facets, which is measured to be 6$\pm$0.5 dB/facet. This is accomplished using a cut-back method by comparing the insertion losses of several devices with similar waveguide geometry but different total lengths. The coupling loss could be reduced by employing mode-converters, such as tapered waveguides, to improve the overlap between the fiber and waveguide modes \cite{mitomi_design_1994}. Finally, we estimate the total on-chip loss including the splitter to be $\sim$ 4 dB and the waveguide propagation loss to be $\sim$ 1 dB, which is comparable with our previous work \cite{wang_integrated_2018}. The small waveguide propagation loss allows advanced architectures such as cascaded modulators to be added without induce a large increase in insertion loss \cite{otsuji_10-80-gb/s_1996}.

One of the most useful figures of merit for EO phase modulators is the frequency dependency of RF V$_\pi$ \cite{gopalakrishnan_performance_1994}. We measure it by driving our phase modulator using sinusoidal electrical signals from an RF signal generator (Hittite HMC-T2240, 0 to 40 GHz). The modulated optical signal is analyzed by an optical spectrum analyzer (OSA) (Yokogawa AQ6370). Fig. \ref{fig2} (a) shows the measured spectra when driving the modulator at RF frequencies from 5 GHz to 40 GHz, where a pair of symmetrical sidebands are generated in the frequency domain \cite{kawanishi_high-speed_2007,shi_high-speed_2003}. We note that reflections between the end-facets of the waveguide give rise to residual intensity modulation of 6\% (of total transmitted optical power). This was measured by driving a modulator with a 100-kHz triangular wave with a voltage exceeding V$_\pi$. This intensity modulation, however quickly drops off at GHz frequencies due to the lifetime ($\sim$ 300 ps) of the Faber-Perot resonance formed between the waveguide facets and velocity mismatch between the RF signal and the resonant standing wave. Therefore, we conclude that sideband generation that we observe is mostly due the phase modulation. RF signals below 5 GHz are not measured since the generated sidebands could not be well separated from the carrier light, due to the limited spectral resolution of the OSA. The EO roll-off of the first sidebands is estimated to be 2.1 dB in the 5 to 40 GHz range. This value was obtained by subtracting the RF cable losses (inset of Fig. \ref{fig2}(a)) from the measured data, indicating a flat frequency response of RF V$_\pi$ over a wide frequency range. For increased RF driving powers, high-order sidebands are generated (Fig. \ref{fig2}(b)).

\begin{figure}
	\centering
	\includegraphics[angle=0,width=0.45\textwidth]{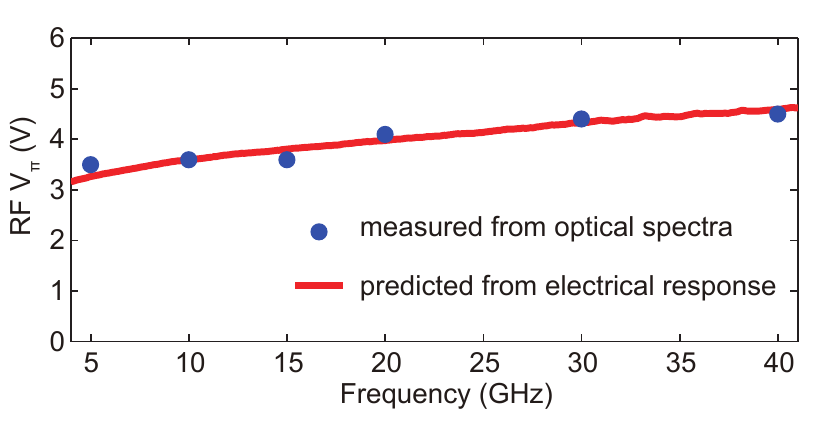}
	
	\caption{\label{fig3}\textbf{Phase modulator RF V$_\pi$ measurements}  Blue dots show the experimental RF V$_\pi$ values from 5 to 40 GHz obtained from the modulated optical spectra. Red curve shows the theoretical RF V$_\pi$ values predicted from the electrical response of the RF electrodes.}
\end{figure}

The power ratio between the modulated sidebands and the carrier band from the measured spectra can be used to determine RF V$_\pi$ at any given frequency. In particular, the powers in the carrier band and the generated sidebands are related via the Bessel function, based on which, we can evaluate the RF V$_\pi$ values using 

\begin{equation}
\textrm{RF} V_\pi=\frac{\pi}{2m}V_\textrm{pp}=\frac{\pi}{m}\sqrt{2PR}
\end{equation}

where $V_\textrm{pp}$ is peak-to-peak drive voltage, $P$ is the power of the RF signal, $R = 50~\Omega$ is the impedance of driving circuit, and $m$ is the measured modulation amplitude, which is the phase change (zero-to-peak) of the phase modulator \cite{kawanishi_high-speed_2007,shi_high-speed_2003}. Fig. \ref{fig3} shows the extracted RF V$_\pi$ values at different frequencies. We find that RF $V_\pi$ only increases slowly, ranging between 3.5 V to 4.5 V in the range of 5 to 40 GHz, which corresponds to an increase of only 28\% over a very broad spectral range. The relative EO roll-off is 2.2 dB, which is consistent with the roll-off calculated from the first sidebands powers (Fig. \ref{fig2}(a)).

We further confirm the measured EO roll-off by calculating the RF V$_\pi$ values from the measured electrical responses (transmission coefficient, $S_{21}$) of the RF electrodes. Assuming perfect velocity matching between RF and optical fields, the dominant origin of EO roll-off arises from the performances of the RF transmission line. The extracted RF V$_\pi$ using this method is in excellent agreement with the optical measurement (Fig. \ref{fig3}), confirming that the RF and optical velocities are well matched. Note that all V$_\pi$ values cited in this work correspond to a $\pi$-phase shift in a single optical path, therefore being roughly twice compared with the Mach-Zehnder interferometer architecture which can employ a push-pull configuration \cite{wang_integrated_2018}. 

In general, V$_\pi$ is directly proportional to the width of the gap and inversely proportional to the length of the electrode. Therefore, smaller gaps and longer electrodes will give rise to a lower V$_\pi$. However, reduced electrode gap increases optical loss due to metal absorption, while longer electrodes contribute to larger microwave loss, and smaller bandwidth. Importantly, even with this trade-off taken into account, our modulator’s performance exceeds that of the commercial counterparts in both optical losses and V$_\pi$ with an overall shorter device length due to the high confinement of the optical mode \cite{thorlabs_40_nodate,eospace_low-loss_nodate,ixblue_lithium_nodate}. Moreover, our RF V$_\pi$ is also flat over a large frequency band due to the reduced metal electrode length.  

\begin{figure}
	\centering
	\includegraphics[angle=0,width=0.45\textwidth]{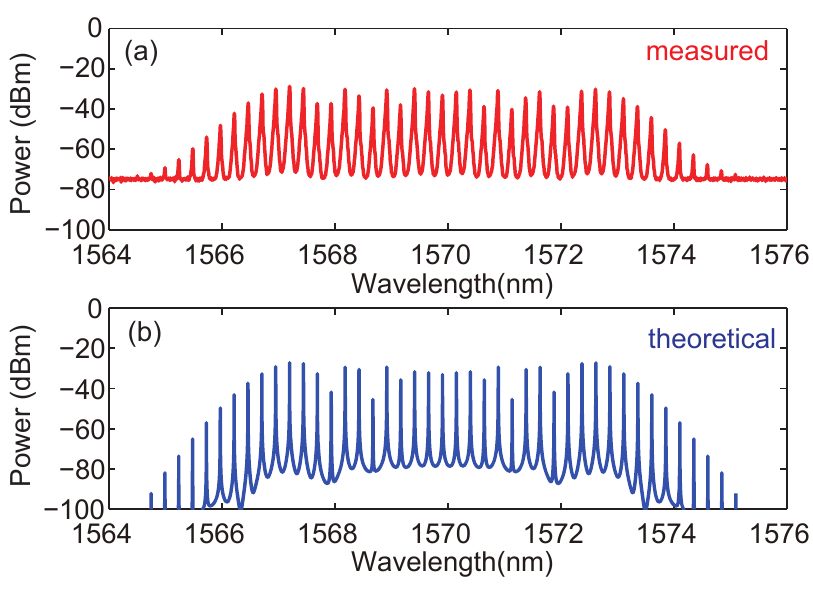}
	\caption{\label{fig4}\textbf{High RF power phase modulation} (a) Measured and (b) theoretical optical spectra of our phase modulator when driven at RF power of 3.1 W at 30 GHz (equivalent to $\sim$ 4 V$_\pi$).}
\end{figure}

One particularly important application of phase modulators is EO comb generation, which is of interest for telecommunication \cite{kippenberg_microresonator-based_2011,zhang_broadband_2018}, ultra-fast optics \cite{gohle_frequency_2005}, and so on. The main challenge to realize this is the requirement of a low RF V$_\pi$ and high RF power handling ability. For this reason, past realizations of EO combs by a single phase modulator are relatively narrow \cite{sakamoto_optoelectronic_2006}. Alternative configurations that can generate broader combs use multiple modulators (commonly three or four) \cite{papp_microresonator_2014}, which requires multiple high power RF drives and introduces additional optical insertion loss. Due to the low RF V$_\pi$ of our phase modulator, we are able to generate a broad EO comb with over 40 spectral lines spanning 10 nm in frequency domain, with a single phase modulator. This is accomplished by driving the modulator by a 30 GHz RF signal amplified to 3.1 W, corresponding to a total acquired phase shift of $\sim$ 4$\pi$. The measured optical spectrum (Fig. \ref{fig4}(a)) is in good agreement with the theoretical prediction (Fig. \ref{fig4}(b)) further confirming the capability of our devices to handle high RF powers without introducing significant signal distortions. For comparison, using modeling we estimate that the same RF power applied to a commercial phase modulator (model LN27S-FC with V$_\pi$ of 7.5 V at 30 GHz) \cite{thorlabs_40_nodate} would result in $\sim$ 6 nm wide comb, with 28 lines. This clearly illustrates the positive role of lower V$_\pi$ on EO comb bandwidth.

In conclusion, we have demonstrated a high-performance phase modulator based on LNOI platform, featuring low on-chip optical losses and low half-wave voltages. The measured RF V$_\pi$ values remain relatively constant over a broad frequency range. The  flatness of the frequency response indicates that the device could also perform well at very high frequencies exceeding 100 GHz \cite{wang_integrated_2018,mercante_thin_2018}, showing high potential for millimeter-wave and terahertz applications. Further reducing the electrode gap and the RF loss could lead to even lower RF V$_\pi$ and broader EO combs. Many phase modulator applications require the modulators to be driven at voltages far exceeding the half-wave voltage \cite{wooten_review_2000}. These applications are consequently limited by the required high voltages and low power-handling ability, which we overcome with the here-presented modulators. Our work shows that phase modulators made from thin-film LN outperform conventional bulk modulators simultaneously in half-wave voltage, optical propagation loss, size and EO bandwidth. Our work therefore enables new ventures in EO signal processing, as well as microwave and terahertz photonics. 

We acknowledge Collins Aerospace, National Science Foundation (NSF) (ECCS-1740296), DARPA SCOUT program (W31P4Q-15-1-0013) and City University of Hong Kong (Start-up Funds) for partly funding this research. We thank R. Morandotti for providing an RF amplifier. Device fabrication was performed at the center for nanoscale systems (CNS) at Harvard University (1541959).

\bibliography{reference}

\end{document}